\documentclass[slac_one]{revtex4}
\usepackage{graphicx}
\usepackage{fancyhdr}
\begin{document}

\title{{\small{2005 International Linear Collider Workshop - Stanford,
U.S.A.}}\\ 
\vspace{12pt}
Monolithic CMOS Pixel R\&D
for the ILC at LBNL} 

%

\author{M.~Battaglia}
\affiliation{Department of Physics, University of California, Berkeley  and\\ 
Lawrence Berkeley National Laboratory, CA 94720, USA}

\author{G.~Abrams, P.~Denes, L.C.~Greiner, B.~Hooberman, L.~Tompkins,
H.H.~Wieman}
\affiliation{Lawrence Berkeley National Laboratory, CA 94720, USA}

\begin{abstract}
An R\&D program on monolithic CMOS pixel sensors for application at 
the ILC has been started at LBNL. This program profits of significant 
synergies with other R\&D activities on CMOS pixel sensors. The project 
activities after the first semester of the R\&D program are reviewed.
\end{abstract}

\maketitle

\thispagestyle{fancy}


\section{INTRODUCTION}

The anticipated ILC physics program requires pixel sensors with performances well beyond 
those obtained for the experiments at LHC in terms of single point resolution and 
material budget. At the same time the ILC experimental environment, with its lower event 
rate and radiation flux, admits Si sensors that are substantially thinner, more precise 
and more segmented than at the LHC and thus motivates new directions for R\&D to 
achieve these performance~\cite{ilc}. 
LBNL has played a major role in Si detector development and 
it is currently engaged in a number of related projects, including the ATLAS pixel project, 
thick CCD development for SNAP, CMOS pixel R\&D for the STAR upgrade and also CPCCDs and CMOS 
pixels for application at synchrotron light sources and electron microscopy. Monolithic CMOS
pixel sensors offer several desirable features for application in collider experiments. The 
STAR experiment at RHIC has engaged in the first vertex detector based on CMOS pixels, under
the guidance of the STAR group at LBNL. The ILC R\&D program aims at addressing key R\&D issues, 
building on the significant know how established at LBNL, and exploiting synergies and 
opportunities at the interface between these different fields of applications.

The R\&D program on CMOS pixel sensors for the ILC started in Fall 2004 and it is addressing 
three main issues: the characterization of pixel sensors and the simulation of their 
response in the ILC physics and background environment, the test and characterization of 
thinned pixel sensors and the design and testing of new pixel test structures embedding 
on chip data reduction capabilities to specifications set by the ILC requirements. This 
paper reviews the project activities after the first semester of the R\&D program.  

\section{SENSOR CHARACTERIZATION}

In order to perform tests of charge collection a setup based on an highly collimated 
light beam has been developed. The system uses a set of LED and laser diodes of 
different wavelengths, in the range 850~nm $< \lambda <$ 1300~nm, pigtailed to a 
single-mode Corning HI1060 optical fiber with a 6~$\mu$m core diameter and a numerical 
aperture of 0.14. The light is focused through an achromatic doublet lens, with 5~mm 
focal length, onto the detector surface. The choice of laser wavelength allows to probe 
the charge collection mechanism from 
a penetration depth of 15~$\mu$m to the full wafer thickness. The system is mounted on a 
pneumatic isolation workstation to reduce the effect of vibrations. The light beam position 
on the detector plane and the focal distance is controlled by a computer-controlled 
$x-y$ stage with sub-$\mu$m resolution.

The laser diode can be operated both in DC and in pulsed mode. Laser pulses are 
generated by modulating a DC current source with a fast pulse through a bias-tee
(Picosecond Pulse Lab Model 5550B). 
The drive pulse is generated by a Picosecond Pulse Lab pulser (Model 2600) and allows 
pulse lengths in the range 0.5-100~ns which can be triggered from the FPGA controlling 
the chip readout, thus ensuring its synchronization with the detector readout cycle. 
In order to reduce the sensor noise, measurements can be conducted in a test chamber able 
to reach $-75^o C$. 

\begin{figure}[t]
\centering
\includegraphics[width=75mm]{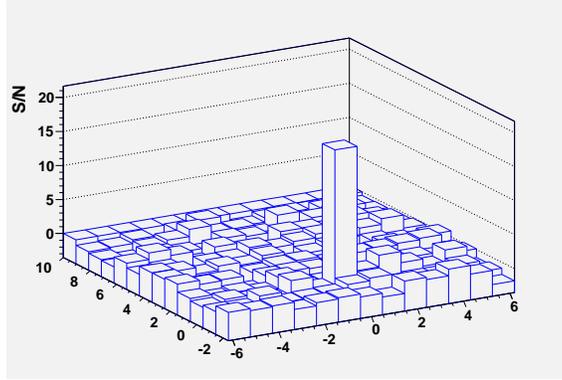}
\caption{Single pixel cluster generated by a LED pulse focused on a 
{\tt MIMOSA-5} pixel detector.} \label{fig:evt}
\end{figure}

The set-up has been tested using several {\tt MIMOSA-5} chips~\cite{mimosa}. 
Light can be focused to a spot smaller than the pitch size of 17~$\mu$m, such that the induced signal 
appears on a single pixel (see Figure~\ref{fig:evt}).
In these conditions, extensive scans can be performed to determine the cluster size and the 
charge collection properties as a function of the position of the light spot to the reference 
pixel.

In the test set-up the detector is mounted on a mezzanine card, specific to the chip design,
connected to a readout board. This includes a Xilinx FPGA for controlling the 
readout sequence and 12-bit ADCs. Data is transferred from the board to the DAQ PC through a 
National Instruments I/O card.
Data reduction can be performed online within a {\tt LabView} program. This program controls 
also the $x-y$ stage position, the laser and the chip bias voltage and current. It computes 
pedestals and noise, in absence of signal, and performs correlated 
double sampling (CDS) and pedestal subtraction. The data is then transformed into the 
{\tt LCIO} format, which is the standard as persistency model for ILC data, and written to disk 
for the subsequent offline analysis. A C++ program performs pedestal and noise initialization, 
if not already done on-line, updates their values through the run and produces tuples for 
further analysis and cluster visualization in {\tt ROOT}. Chip pedestal, noise and dead or noisy
pixels can also be stored in a condition database and retrieved for analysis. The use of a 
standard {\tt LCIO} format, allows the same clustering and reconstruction software to be 
applied to both real lab test data and events simulated using {\tt GEANT 4}.

\section{SENSOR BACK-THINNING}

Sensor thickness is a major issue in the design of the ILC Vertex Tracker. The 
requirement to obtain a multiple scattering term as low as $\simeq 10$~($\mu$m/GeV)~$/p_t$ 
in the track extrapolation accuracy, for a multi-layered detector, pushes on the sensor 
thickness. Further, the amount of material 
located in front of the calorimeter needs to be minimized to allow for optimal energy flow 
performance. CMOS pixel sensors, with their active epitaxial layer of only few $\mu$m to 
$\simeq 15$~$\mu$m, can be back-thinned without losing, in principle, any significant fraction 
of the collected signal. However, there are several factors which need to be verified. 
Signal losses may arise both from disturbances to the charge collection mechanism introduced 
by the back-thinning process as well as by the loss of charge carriers created in the bulk. 
Several {\tt MIMOSA-5} chips have already been back-thinned to 50~$\mu$m~\cite{aptek}, tested 
with a $^{55}Fe$ source and found to function. The thinning has been performed on diced chips
with an yield of $\simeq$94\% and a thickness constant to better than 5~$\mu$m. 
In order to carefully assess 
the effects of the back-thinning process on the charge collection mechanism, a new set of tests 
is presently being conducted. They aim at a full characterization of the same pixel sensors, 
both before and after back-thinning, for various thickness values. A set of six {\tt MIMOSA-5} 
chips are being used for these tests. The test procedures includes the characterization of the 
chip response to laser light of different wavelengths and to a collimated $^{55}Fe$ X-ray source.
The procedure being adopted is as follows.
Detectors are first mounted on a bare mezzanine card using WaferGrip dissolvable thin film 
adhesive and wire-bonded. Afterward the detector chip is masked and the mezzanine card is 
loaded with its components. After having been fully characterized in the test setup described 
above, the detector is removed and sent for back-thinning~\cite{aptek}. The thinned detector is 
then glued back on the mezzanine card and wire bonded. This procedure has already been 
successfully tested. The back-grinding process is able to accommodate for the 
small residue of bond wire left on the bonding pads and deliver the chips ground to the correct 
thickness and with parallel sides. We begin by characterizing the detector gain by calibrating to the 
$^{55}Fe$ 5.9~keV peak.  Since the gain depends on the reference voltage supplied to each of the 4 sectors 
of the detector this also allows to optimize the voltage setting. The study of the charge collected for the 
5.9~keV X-rays converting under a single pixel, will provide a first characterization of 
the charge collection properties of the sensors before and after thinning. The detector response to the 
collimated laser spot, will allow to compare the charge collection process for charge carriers produced 
at different depths, depending on the laser wavelength.

\section{CHIP DESIGN}

\begin{figure}[bh!]
\centering
\includegraphics[width=75mm]{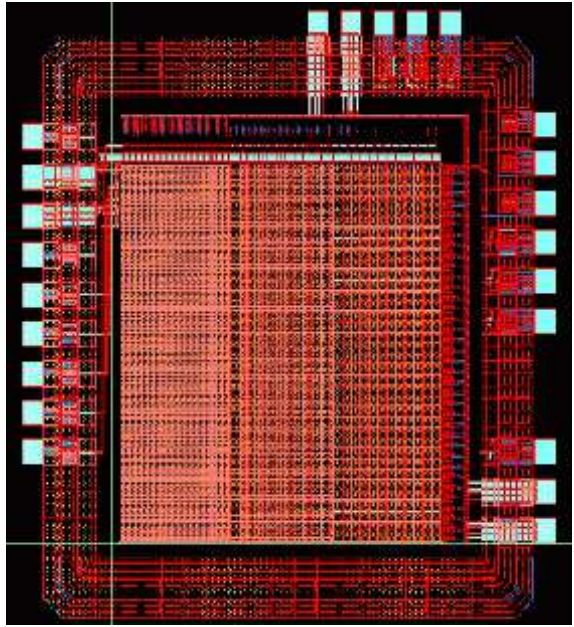} 
\caption{Layout of the CMOS pixel chip designed at LBNL and produced in the AMS 0.35 OPTO process. 
The chip has 48$\times$144 10~$\mu$m pixels, 24$\times$72 20~$\mu$m pixels and 
12$\times$36 40~$\mu$ pixels.} \label{fig:chip}
\end{figure}

We are leveraging developments in other areas in order to support vertex detector R\&D for the ILC.
As a first step, we have made a small active pixel test chip in the AMS 0.35~$\mu$m OPTO process.  
Processes specifically for CMOS imagers (so-called OPTO) are excellent candidates for particle 
detection applications, as they have thicker epitaxial regions, and thus better signal to noise 
performance, and much lower leakage current compared to conventional digital processes.  The circuit 
consists of arrays of 3-transistor active pixels.  An identical 3T layout is repeated at 10, 20 and 
40~$\mu$m pitch in order to study charge collection properties and optimize the pixel size 
(see Figure~\ref{fig:chip}).  Five analog outputs are provided, and the column multiplexers are 
arranged so that five adjacent pixels are output at the same time. Testing of the chip will be carried 
out in Fall 2005.  

Although this first test circuit has only analog outputs, a device for the ILC should have on-chip 
digitization. We have begun to study ADC options. One possibility is to use a per-column Wilkinson 
architecture.  Active pixel sensors in 0.25~$\mu$m CMOS have already been developed at LBNL, for a 
different application, with 10-bit Wilkinson ADCs on a 19~$\mu$m pitch.  Although the Wilkinson 
architecture is attractive, as it is simple and has a low power dissipation, speed capabilities are 
marginal for the ILC application. Therefore we are presently considering a pipelined approach for 
a pixel architecture with analog CDS.
An N-bit pipelined ADC can, in principle, be constructed from N cells consisting of a 1-bit ADC, 
a 1-bit DAC and a gain 2 amplifier.  As the operation is pipelined, each stage can 
make use of the full digitizing time, i.e. a 4-bit pipelined ADC running at 50~MHz would provide 
a digital 
output every 20~ns, with a 60~ns latency and each cell would use the full 20 ns for the 1-bit operation.  
The 1-bit approach fails when offsets are taken into account, so 1.5-bit ADCs and DACs are used.  
As precision requirements are reduced, the pipelined approach improves in speed, power consumption and 
size, so that a 4-5 bit pipelined ADC on a 20~$\mu$m pitch at $\simeq$1~mW/channel should be feasible.

\begin{acknowledgments}
This work was supported by the Director, Office of Science, of the U.S. Department 
of Energy under Contract No. DE-AC02-05CH11231.
This research used resources of the National Energy Research Scientific 
Computing Center, which is supported by the Office of Science of the U.S. 
Department of Energy under Contract No. DE-AC03-76SF00098.
This activity is carried out in collaboration with IReS, Strasbourg (France). 
We wish to thank Marc Winter for sharing with us precious know how and providing 
several chips developed by his group. We are also grateful to Howard~Matis and 
Michelle~Tuchscher for their help.
\end{acknowledgments}


\begin{thebibliography}{99}

\bibitem{ilc}
M.~Battaglia, Nucl. Instr. and Meth. {\bf A 530} (2004), 33.

\bibitem{mimosa}
Yu. Gornushkin {\it et al.}, Nucl. Instr. and Meth. {\bf A 513} (2003), 291. 

\bibitem{aptek} The back-thinning is a commercial process done at Aptek Industries, Inc., 
San Jose, CA (http://www.aptekindustries.com/).

\end{thebibliography}
\end{document}